\documentclass{ws-ijmpd}
\usepackage{url}
\usepackage[super,compress]{cite}
\usepackage[usenames,dvipsnames]{color}
\usepackage{epstopdf}
\usepackage[linktocpage]{hyperref}
\usepackage{cleveref}
\definecolor{coolblack}{rgb}{0.0, 0.18, 0.39}
\definecolor{darkred}{rgb}{0.5,0,0}
\definecolor{darkgreen}{rgb}{0,0.5,0}
\definecolor{darkblue}{rgb}{0,0,0.5}
\definecolor{lapislazuli}{rgb}{0.15, 0.38, 0.61}
\definecolor{venetianred}{rgb}{0.78, 0.03, 0.08}
\definecolor{bleudefrance}{rgb}{0.19, 0.55, 0.91}
\definecolor{dogwoodrose}{rgb}{0.84, 0.09, 0.41}
\definecolor{dogwoodrose}{rgb}{0.84, 0.09, 0.41}
\hypersetup{colorlinks=true, citecolor=darkgreen, linkcolor=darkblue, 
urlcolor = venetianred}

\usepackage[utf8]{inputenc}
\usepackage[T1]{fontenc}

\newcommand{\esdeo}[1]{\textcolor{green}{#1}}

\def\be{\begin{equation}}
\def\ee{\end{equation}}
\def\bea{\begin{eqnarray}}
\def\eea{\end{eqnarray}}
\def\l{\left}
\def\r{\right}

\begin{document}

\markboth{C. F. B. Macedo et al.}
{Scalar waves in regular Bardeen black holes: Scattering, absorption and quasinormal modes}

%
\catchline{}{}{}{}{}
%

\title{Scalar waves in regular Bardeen black holes: Scattering, absorption and quasinormal modes}

\author{Caio F. B. Macedo\footnote{caio.macedo@tecnico.ulisboa.pt}} 
\address{CENTRA, Departamento de F\'isica, Instituto Superior T\'ecnico, Universidade de Lisboa,
Avenida Rovisco Pais 1, 1049 Lisboa, Portugal}

\author{Lu\'is C. B. Crispino\footnote{crispino@ufpa.br}~ and~ Ednilton S. de Oliveira\footnote{ednilton@pq.cnpq.br}}
\address{Faculdade de F\'{\i}sica, Universidade Federal do Par\'a, 66075-110, Bel\'em, Par\'a, Brazil}

\maketitle


\begin{abstract}
We discuss the phenomenology of massless scalar fields around a regular Bardeen black hole, namely absorption cross section, scattering cross section and quasinormal modes. We compare the Bardeen and Reissner-Nordstr\"om black holes, showing limiting cases for which their properties are similar.

\end{abstract}

\keywords{wave scattering; regular black holes}

\ccode{PACS numbers:
04.70.-s, 
04.70.Bw, 
11.80.-m, 
04.30.Nk 
}


\section{Introduction}	

Since the very first years of General Relativity (GR), curvature singularities were found in the field solutions. In fact, the first solution describing a spherical self-gravitating body was found to be unavoidably singular if its radius is smaller than $2M$, where  $M$ is the total mass of the body\footnote{Throughout this work, we use units such that $G=c=\hbar=1$.}. These fully collapsed objects are called black holes (BHs) and they are the most plausible candidates for the supermassive objects in the center of the galaxies~\cite{Narayan:2005ie}.

The presence of singularities in BH spacetimes seems to be a generic feature of GR. In fact, singularity theorems prove that, from some generic hypothesis in the collapsing matter, curvature singularities would be unavoidable in BH formation\cite{Penrose:1964wq,Penrose:1969pc,hawkingb}. Notwithstanding, modifications of GR at quantum level are expected to remove these singular points due to changes of the theory at Planck scale~\cite{Clifton:2011jh}.

One natural question that arises when searching for BHs that present no singularities is whether we can achieve them by relaxing the matter hypothesis in the singularity theorems. One of the first regular BHs, found by Bardeen,\cite{bardeen} in fact violates one of the energy conditions (namely the weak energy condition). The Bardeen BH can be though to be a solution of a nonlinear electrodynamics model with a magnetic monopole\cite{AyonBeato:2000zs}. After Bardeen's proposal, many regular BHs motivated by nonlinear electrodynamics appeared in the literature\cite{Ansoldi:2008jw}.

Building up on recent
investigations~\cite{Macedo:2015qma,Macedo:2014uga,PhysRevD.86.064039}, 
in this work we revisit the problem of scalar planar waves impinging upon a Bardeen BH,
exhibiting complementary numerical results 
for the absorption and (differential) scattering cross sections. 
We also point out some characteristics regarding the quasinormal modes of Bardeen BHs in the eikonal limit.

The Bardeen BH is described by the following line element
\be
ds^2=-f_{\text{BD}}(r)\,dt^2+f_{\text{BD}}(r)^{-1}dr^2+r^2(d\theta^2+\sin^2\theta d\varphi^2),
\label{eq:ds}
\ee
where the lapse function $f_{\text{BD}}(r)$ is given by
\be
f_{\text{BD}}(r)=1-\frac{2Mr^2}{(r^2+q_{\text{BD}}^2)^{3/2}}.
\ee
The Bardeen BH presents two horizons, similarly to the Reissner-Nordstr\"om (RN) BH, whose line element
differs formally from the Bardeen BH one by the lapse function, namely:
\be
f_{\text{RN}}(r)=1-\frac{2M}{r}+\frac{q_{\text{RN}}^2}{r^2}.
\ee
Here $q_{\text{BD}}$ is the magnetic monopole charge of the Bardeen BH 
and $q_{\text{RN}}$ is the electric charge of the RN BH.

In order to better compare Bardeen and RN situations, we shall work with the normalized charges, defined as
$Q_i \equiv q_i/q_{\text{ext}}$, 
where $i = {\text{BD}}, {\text{RN}}$ and
$q_{\text{ext}}$ is the extreme charge in each case~\cite{Macedo:2014uga}.

The Bardeen BH has basically the same causal structure {of} the RN BH\cite{Ansoldi:2008jw}. Additionally, their absorption and scattering cross sections are similar in certain regimes, as we shall point out in this work. 
Moreover, the quasinormal frequencies of the Bardeen and RN spacetimes can be equivalent for high multipolar modes.

\section{Wave analysis and approximations}
\label{sec:dev}
When one considers the scalar wave phenomenology in BH spacetimes, normally one seeks to reduce the problem to a one-dimensional Schr\"odinger-like equation. For the regular Bardeen BH, the scenario is essentially the same. The wave function $\Phi(t,r,\theta,\varphi)$ for a given frequency $\omega$ can be described by
\be
\Phi=\frac{\psi(r)}{r}Y^{m}_{l}(\theta,\varphi)e^{-i\omega t},
\ee
where $Y^{m}_{l}$ are the spherical harmonics. The function $\psi(r)$ obeys the one-dimensional Schr\"odinger-like equation
\be
\l(-\frac{d}{dr_*^2}+V_\phi(r) -\omega^2\r)\phi(r)=0,\label{eq:eqr}
\ee
{where we have defined the tortoise coordinate $d/dr_* \equiv f\,d/dr$ and the } potential $V_\phi$ is given by
\be
V_\phi(r)=f\left(\frac{l (l+1)}{r^2}+\frac{f'}{r}\right).\label{eq:pot}
\ee
{Incident planar waves can be described by the so called $in$ modes, which present the following asymptotic behavior}
\be
\phi(r)\sim\l\{
\begin{array}{ll}
\mathcal{A}_{\omega l} R_I+\mathcal{R}_{\omega l}R_I^{*},&{\rm as}~ r_*\to +\infty ~(r\to +\infty),\\
  \mathcal{T}_{\omega l} R_{II}, &{\rm as}~r_*\to - \infty ~(r\to r_h),
\end{array}\r.
\label{eq:inmodes}
\ee
with
\be
R_I=e^{-i \omega r_*}\sum_{j=0}^N \frac{A^{(j)}_\infty}{r^j},~R_{II}=e^{-i \omega r_*}\sum_{j=0}^N (r-r_h)^j A^{(j)}_{r_h},\label{eq:rii}
\ee
where $A^{(0)}_{\infty,r_h}=1$, and $|\mathcal{R}_{\omega l}|^2$ and $|\mathcal{T}_{\omega l}|^2$ are the reflection
and transmission coefficients, respectively. 
Quasinormal modes can also be found from Eq.~(\ref{eq:inmodes}), 
with the additional restriction of no incoming waves from the past infinity, 
i.e., $\mathcal{A}_{\omega l}=0$, what is associated with a boundary value problem, 
with the eigenvalues being the quasinormal mode frequencies $\omega_{n}$.

When investigating the scattering of fields, 
the key point is to find the differential scattering cross section. 
Some standard procedures can be considered for such purpose, 
like $(i)$ the geodesic analysis, 
$(ii)$ the semiclassical 
approach, and 
$(iii)$ the full wave analysis. 
The geodesic analysis is an important approximation because at very high frequencies the 
wave propagates along null geodesics\cite{Collins:1973xf}. 
The semiclassical approach captures the wave interferences, 
although it fails to describe the differential scattering cross section for low angles\cite{Matzner:1968}. 
Finally, the planar wave analysis describes the full phenomenology, 
with the drawback of having to solve the full wave equation (usually numerically). 
The key formulas for the differential scattering cross section are
\be
\frac{d\sigma}{d\Omega}=\l\{
\begin{array}{cl}
	\frac{1}{\sin\chi}\sum{b(\chi)\l|\frac{db(\chi)}{d\chi}\r|},& {\rm classical},\\
	2\pi \omega b_g^2 \l|\frac{db}{d\theta}\r|_{\theta = \pi} J_{0}^2(\omega b_g\sin\theta),& {\rm semiclassical},\\
	\l|\frac{1}{2 i \omega} \sum\limits_{l = 0}^{\infty}(2l+1)\left[(-1)^{l+1} \mathcal{R}_{\omega l} - 1\right] P_l(\cos\theta)\r|^2,& {\rm planar~ wave},
\end{array}\r.
\label{scs_approxs}
\ee
where $b(\chi)$ is the impact parameter associated with a
scattering angle $\chi$, $b_g$ is the impact parameter of backscattered waves, 
$J_{0}$ is the Bessel function of the first kind and $P_l$ are the Legendre polynomials.

For the absorption cross section, we also have some standard approaches.
For low frequencies, the absorption cross section 
of stationary BHs equals the area $A_h$ of the BH horizon~\cite{Higuchi:2001si}. 
For mid-to-high frequencies an approach was pointed out in Ref.~\refcite{Decanini:2011xi} 
which captures the important features of the absorption cross section, 
including the geodesic limit for higher frequencies. 
This approach is known as $sinc$ approximation.
These two results
can be used as approximations to the full partial wave analysis, 
which gives correct outputs
for the whole frequency range (within numerical approximation). 
We have
\be
\sigma_{\rm abs}=\l\{
\begin{array}{cl}
  A_{h}\equiv 4\pi {r_{h}}^2, & \omega\sim 0,\\
	\pi b_c^2-4 \pi \frac{\lambda b_c^2}{\omega}e^{-\pi\lambda b_c}\sin\l(2\pi\omega b_c\r),& sinc\\
	\sum\limits_{l=0}^{\infty}\frac{\pi}{\omega^2}(2l+1)\l|\mathcal{T}_{\omega l}\r|^2, &{\rm planar~ wave},
\end{array}\r.
\label{acs_approxs}
\ee
where $r_h$ is the horizon radius, 
$b_c$ is the critical impact parameter for null geodesics 
and $\lambda$ is the Lyapunov exponent of null circular geodesics~\cite{Cardoso:2008bp}.

In the so-called eikonal approximation (high $\omega$ and $l$) the wave propagates as a null geodesic.
In this limit, high multipole quasinormal modes are well approximated by~\cite{Cardoso:2008bp}
\be
\omega_n \sim b_c^{-1}-i(n+1/2)\l|\lambda\r|.
\label{wn}
\ee

From the above expressions, we can draw some conclusions 
when we compare Bardeen and RN BHs. 
First, concerning the scattering cross section, 
we see that whenever the Bardeen and RN 
BHs present the same value of $b_g$, 
the pattern of the backscattered waves is the same. 
Second, we see that the absorption of Bardeen and RN 
BHs is similar in the 
mid-to-high frequencies [cf. the sinc expression in Eq.~(\ref{acs_approxs})], 
whenever the critical impact parameter $b_c$ of the two spacetimes is the same. 
Finally, the similarities between the Bardeen and RN spacetimes 
with the same $b_c$ are also shared with the high-multipole quasinormal modes frequencies, 
mainly because its expression [cf. Eq.~(\ref{wn})] 
is written in terms of geodesic quantities. In this sense, 
we expect Bardeen and RN black holes 
with the same $b_c$ to have the same response to perturbations. 
We shall explore the aforementioned characteristics in the next section.

\section{Numerical analysis and comparisons}
\label{numerics}

In this section we shall focus on cases for which the absorption and scattering cross section 
of the Bardeen and RN BHs are similar. 
As mentioned in the previous section, the similarities between the absorption cross section 
of Bardeen and RN BHs can be understood by the same reasons as for the similarities between the quasinormal modes in the eikonal limit. A more detailed analysis of the absorption, scattering and quasinormal modes of regular Bardeen BHs can be found in 
Refs.~\refcite{Macedo:2015qma,Macedo:2014uga,PhysRevD.86.064039,Flachi:2012nv}.

\begin{figure}
\center
\includegraphics[width=0.7\textwidth]{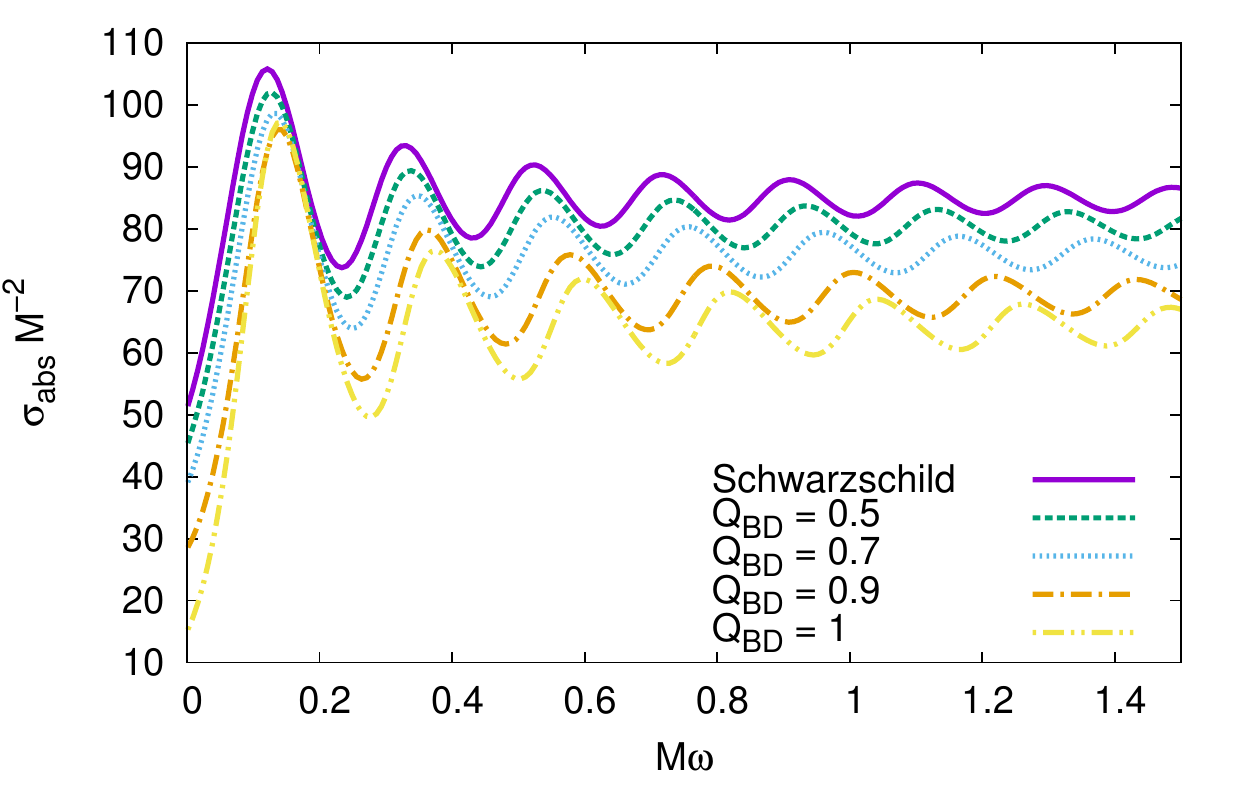}\\
\includegraphics[width=0.7\textwidth]{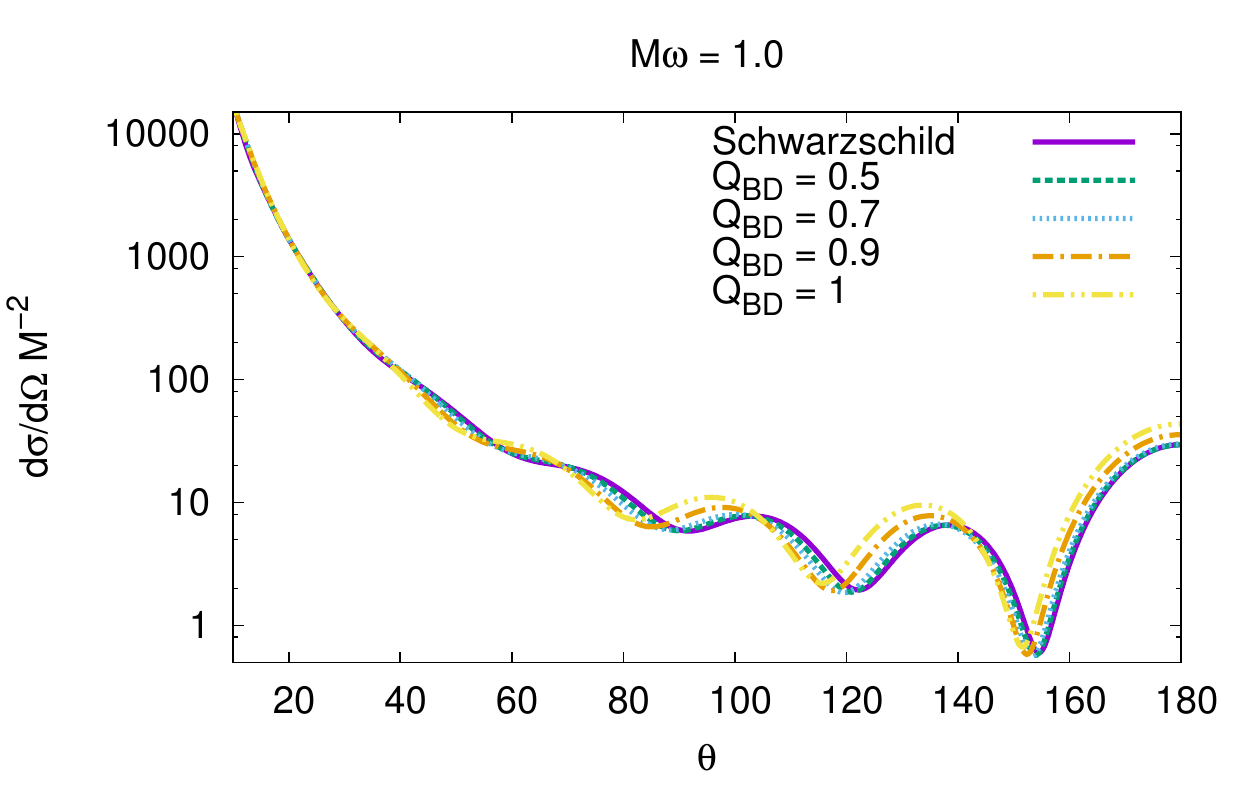}
\caption{Comparison between the absorption (top) and differential scattering (bottom) 
cross sections of Bardeen and Schwarzschild BHs. 
We can see that the Bardeen case is qualitatively similar to the case of BHs with singularities.}%
\label{fig:compsch}%
\end{figure}
In Fig. \ref{fig:compsch} we compare the Bardeen BH absorption and 
(differential)
scattering cross sections with the ones of the Schwarzschild spacetime ($Q_i=0$). 
The Bardeen absorption cross section is always smaller than the one of the Schwarzschild BH, 
for any fixed arbitrary value of the field frequency $\omega$. 
As for the scattering cross section, like in the RN spacetime, 
one of the main influences of the BH charge 
is a shift in the deflected angle~\esdeo{\cite{Crispino2009-prd79_064022}}, 
what can be clearly noticed in the scattering of null rays\cite{Macedo:2015qma}.

\begin{figure}
 \centering
 \includegraphics[width=0.7\textwidth]{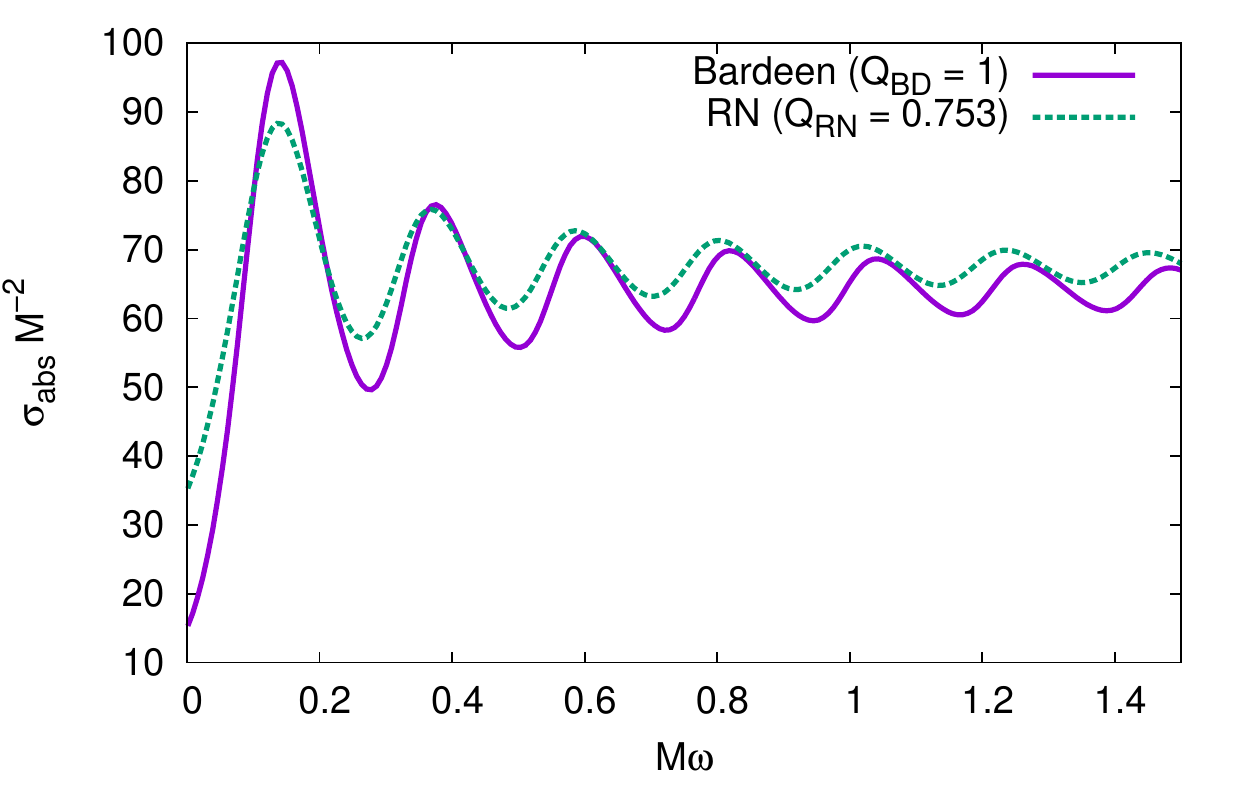}\\
 \includegraphics[width=0.7\textwidth]{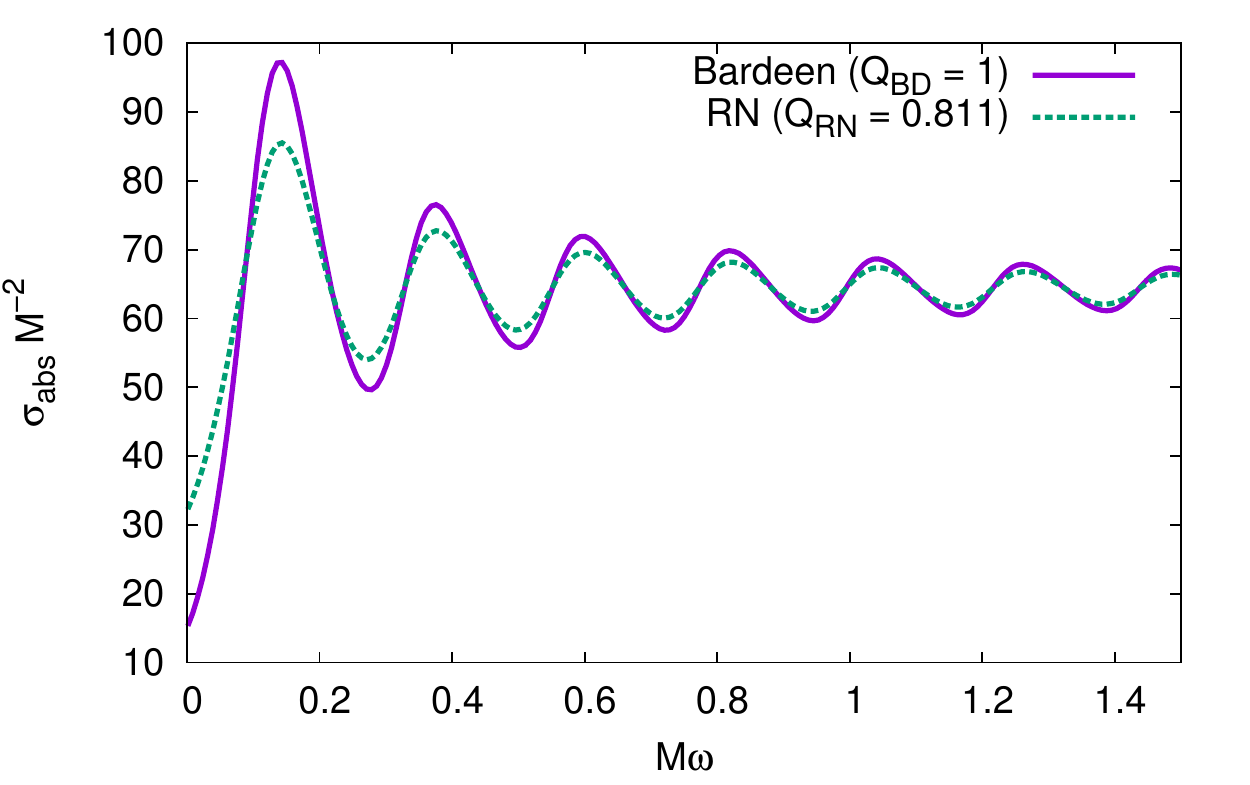}
\caption{Scalar absorption cross section of the extreme Bardeen BH compared with RN BHs 
with $Q_{RN} = 0.753$ (top) and $Q_{RN} = 0.811$ (bottom).}
 \label{fig:abs_comp}
\end{figure}

In Fig.~\ref{fig:abs_comp} we compare the absorption cross section 
of Bardeen BHs with $Q_{BD} = 1$ and RN BHs 
with $Q_{RN} = 0.753$ and $Q_{RN} = 0.811$. 
The value of the RN BH charge $Q_{RN} = 0.753$ ($Q_{RN} = 0.811$) was chosen 
such that the RN BH presents the same value of $b_g$ ($b_c$) of an extreme Bardeen BH. 
The capture cross section of photons 
by extreme Bardeen BHs coincides with the one for RN BHs with $Q_{RN} = 0.811$. 
We see from Fig.~\ref{fig:abs_comp} that even for the cases with coinciding values of $b_g$ (left plot) and $b_c$ (right plot),
the difference  between the overall absorption cross sections of Bardeen and RN BHs can be clearly seen.

\begin{figure}
\center
 \includegraphics[width=0.7\textwidth]{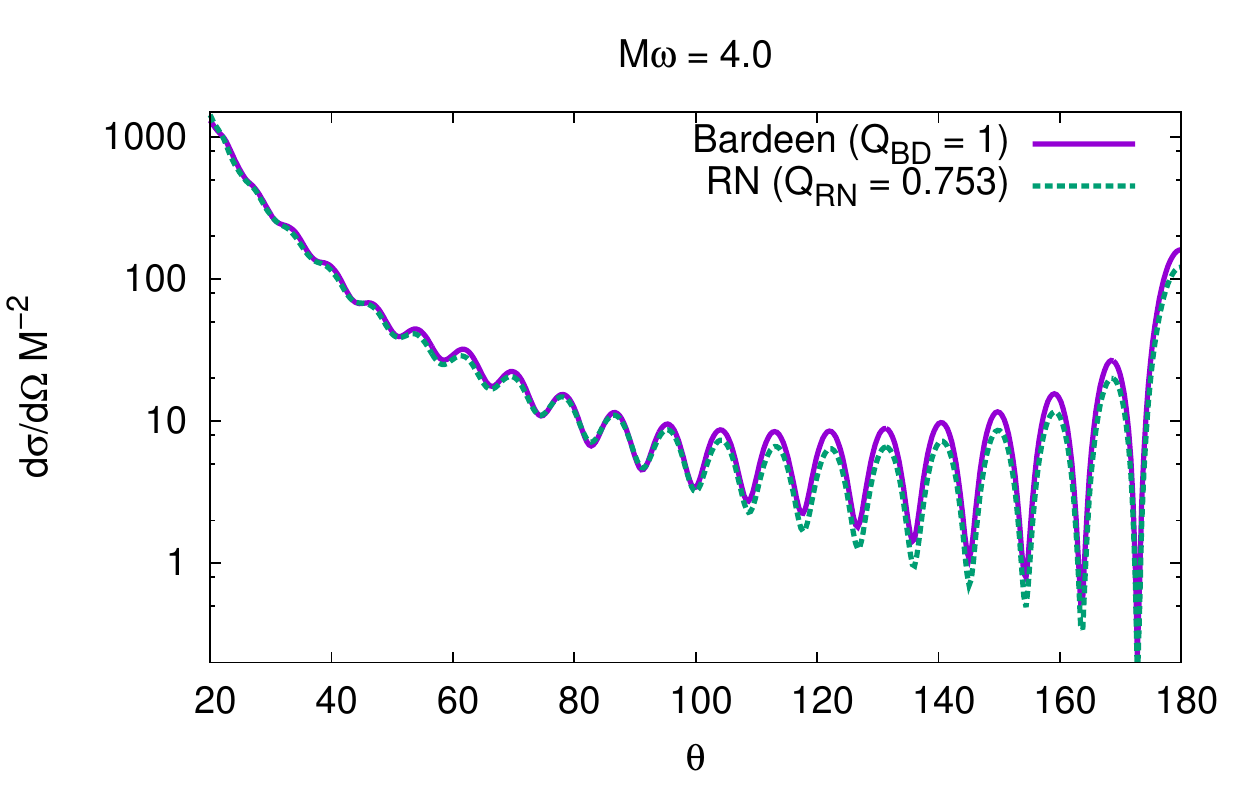}\\
 \includegraphics[width=0.7\textwidth]{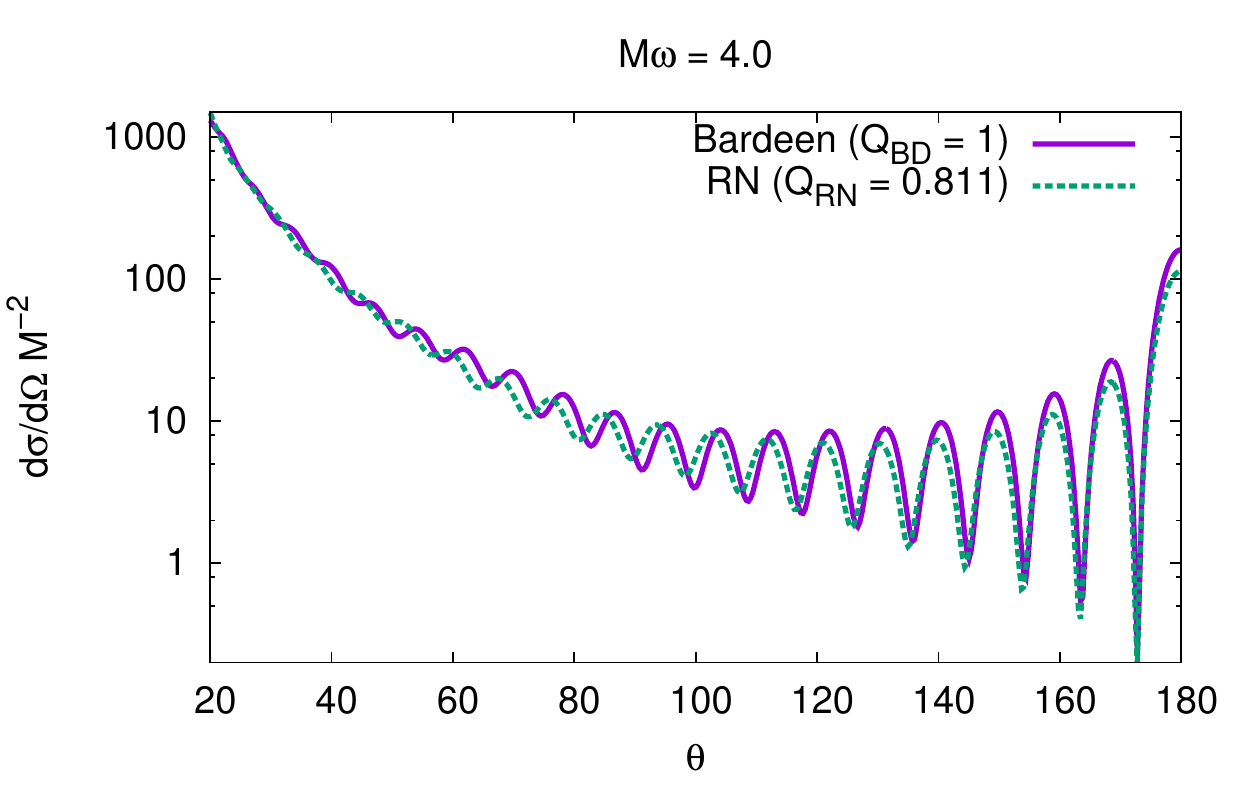}
\caption{Scalar differential scattering cross section of the extreme Bardeen BH compared with RN BHs 
with $Q_{RN} = 0.753$ (top) and $Q_{RN} = 0.811$ (bottom).}
 \label{fig:rn_bd-sca_comp}
\end{figure}

In Fig.~\ref{fig:rn_bd-sca_comp} we show the comparisons of the 
differential scattering cross section of extreme Bardeen 
and RN BHs with charges $Q_{RN} = 0.753, 0.811$. 
Analogously to what happens to the absorption cross section
(cf. Fig.~\ref{fig:abs_comp}),
there is no general coincidence of the differential scattering cross section of 
Bardeen BHs and RN BHs with the same $b_g$ (left plot) or $b_c$ (right plot).

\begin{figure}%
\center
\includegraphics[width=.7\columnwidth]{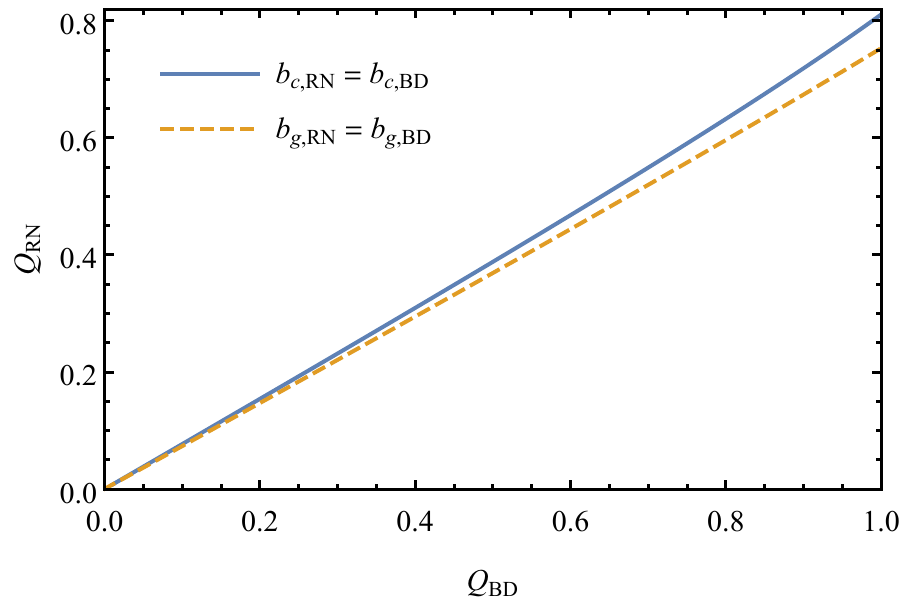}%
\caption{Pair of normalized charges for which the 
Bardeen and RN BHs have the same values of $b_c$ (solid line) and $b_g$ (dashed line). 
For each point of the $b_{c,\rm RN}=b_{c,\rm BD}$ (solid) line 
the absorption spectrum for the two BHs have the closest
profile for mid-to-high frequencies and quasinormal modes, 
while for each point of the $b_{g,\rm RN}=b_{g,\rm BD}$ 
(dashed) line the scattering cross section of the two BHs have 
the closest oscillatory profile.}%
\label{fig:equiv}%
\end{figure}
	
The above features lead us to the conclusion that the oscillatory behavior 
of the absorption and scattering cross sections of 
Bardeen and RN BHs can be similar in certain conditions. 
For the absorption cross section, the major similarity is manifest
for the cases in which the two spacetimes have the same critical impact parameter $b_c$, 
while for the differential scattering cross section it happens 
when the two spacetimes have the same value of $b_g$. 
The value of the charges for which 
$b_{c,\rm RN}=b_{c,\rm BD}$ and $b_{g,\rm RN}=b_{g,\rm BD}$ 
generates lines in the parameters space
of the Bardeen and RN charges. In Fig. \ref{fig:equiv} 
we plot these lines. 
We can see that, for instance, as
previously mentioned, for extreme Bardeen spacetimes, 
the RN BHs that present 
the closest behavior
correspond to $Q_{RN} =0.811$ for the absorption cross section and 
$Q_{RN} =0.753$ for the scattering cross section.

\section{Conclusion}

Absorption and scattering by BHs have been widely studied in the literature
over the past fifty years (cf. Ref.~\refcite{Futterman:1988ni} for references prior to 1988).
Recently, new investigations have been performed, using modern numerical 
and algebraic computational softwares, studying the interaction of different BHs with 
scalar~\cite{Macedo:2013afa, Crispino:2013pya, Benone:2014qaa, bc2016, mlc2016}, 
electromagnetic~\cite{Crispino:2007qw, co2008, cho2009, Crispino2009-prd79_064022, chm2010, cdho2014}, 
spinorial~\cite{Doran:2005vm, Dolan:2006vj}, 
and gravitational fields~\cite{Dolan:2008kf, Oliveira:2011zz, cdho2015}.
Moreover, the absorption~\cite{Crispino:2007zz, Oliveira:2010zzb}
and scattering ~\cite{Dolan:2009zza, doc2011}
by  BH analogues in fluids 
(which do not satisfy Einstein's equations)
have also appeared in the literature, in the last few years.

Concerning BHs in the context of GR, most of the studies have been focused 
in systems presenting singularities hidden inside event horizons.
In this work we have been concerned with some properties of the regular Bardeen BH, 
considering their scalar absorption and scattering cross section, 
as well as some characteristics of the quasinormal modes in the eikonal limit. 
We pointed out some limiting cases for which the Bardeen and 
RN BHs have similar behaviors in the absorption and 
(differential) scattering cross sections, 
as well as in the quasinormal modes in the eikonal limit.

Comparison between RN and Bardeen BHs shows more similarities 
between these two kinds of BHs if $b_{c,\rm RN}=b_{c,\rm BD}$ 
in the case of absorption and quasinormal modes and $b_{g,\rm RN}=b_{g,\rm BD}$ in the case of scattering. 
These similarities can be foreseen from the analytical approximations 
for the absorption and scattering cross sections as well for quasinormal modes. 
For instance, we have seen, from the ``sinc'' approximation in Eq.~\eqref{acs_approxs}, that $b_c$ rules 
both the high-frequency regime as well as the width of oscillations of the absorption cross section. 
For the scattering, $b_g$ is the key quantity in the 
semiclassical ``glory'' approximation, in Eq.~\eqref{scs_approxs},
playing a major role in both interference-fringe widths and scattered-flux intensity.

\section*{Acknowledgments}

The authors would like to thank 
Conselho Nacional de Desenvolvimento Cient\'ifico e Tecnol\'ogico (CNPq), 
Coordena\c{c}\~ao de Aperfei\c{c}oamento de Pessoal de N\'ivel Superior (CAPES), 
and Funda\c{c}\~ao Amaz\^onia de Amparo a Estudos e Pesquisas do Par\'a (FAPESPA), 
from Brazil, for partial financial support.
This work was also supported by the NRHEP 295189 FP7-PEOPLE-2011-IRSES Grant 
and by FCT-Portugal through projects IF/00293/2013 and CERN/FP/123593/2011, 
and by the European Community through the Intra-European Marie Curie Contract No.~AstroGRAphy-2013-623439.

\bibliographystyle{ws-ijmpd}
\bibliography{refs_regular}

\end{document}